\documentclass[showpacs,showkeys,superscriptaddress]{revtex4-2}
\usepackage{amsmath}
\usepackage{amssymb}
\usepackage{graphicx}
\usepackage{xcolor}
\usepackage{ulem}

\begin{document}
\title{Mixed neutron-star-plus-wormhole systems: Rotating configurations
 }

\author{
Vladimir Dzhunushaliev
}
\email{v.dzhunushaliev@gmail.com}
\affiliation{
Department of Theoretical and Nuclear Physics,  Al-Farabi Kazakh National University, Almaty 050040, Kazakhstan
}
\affiliation{
Institute of Experimental and Theoretical Physics,  Al-Farabi Kazakh National University, Almaty 050040, Kazakhstan
}
\affiliation{
Academician J.~Jeenbaev Institute of Physics of the NAS of the Kyrgyz Republic, 265 a, Chui Street, Bishkek 720071, Kyrgyzstan
}
\affiliation{
Institut f\"ur Physik, Universit\"at Oldenburg, Postfach 2503 D-26111 Oldenburg, Germany
}

\author{Vladimir Folomeev}
\email{vfolomeev@mail.ru}
\affiliation{
Institute of Experimental and Theoretical Physics,  Al-Farabi Kazakh National University, Almaty 050040, Kazakhstan
}
\affiliation{
Academician J.~Jeenbaev Institute of Physics of the NAS of the Kyrgyz Republic, 265 a, Chui Street, Bishkek 720071, Kyrgyzstan
}
\affiliation{
Institut f\"ur Physik, Universit\"at Oldenburg, Postfach 2503 D-26111 Oldenburg, Germany
}
\affiliation{
Laboratory for Theoretical Cosmology, International Centre of Gravity and Cosmos,
Tomsk State University of Control Systems and Radioelectronics (TUSUR),
Tomsk 634050, Russia
}

\author{Burkhard Kleihaus}
\email{b.kleihaus@uni-oldenburg.de}
\affiliation{
Institut f\"ur Physik, Universit\"at Oldenburg, Postfach 2503 D-26111 Oldenburg, Germany
}

\author{Jutta Kunz}
\email{jutta.kunz@uni-oldenburg.de}
\affiliation{
Institut f\"ur Physik, Universit\"at Oldenburg, Postfach 2503 D-26111 Oldenburg, Germany
}

\begin{abstract}
We present rapidly rotating neutron stars featuring wormholes in their centers.
They arise in general relativity in the presence of a ghost scalar field.
The nuclear matter is described by a polytropic equation of state, yielding realistic masses and radii for the neutron stars.
The wormholes possess small circumferential radii of size up to 3 km.
With increasing wormhole size, the masses and radii of the stars decrease, while the domain of existence
of these rotating mixed neutron-star-plus-wormhole systems retains the characteristic properties of a rotating neutron star domain.
The question of stability of the mixed configurations under consideration is briefly discussed.
\end{abstract}

\maketitle

\section{Introduction}

Wormholes have been the subject 
of numerous investigations in the past few decades
(see, e.g., Refs.~\cite{Visser:1995cc,Alcubierre:2017pqm}).
To a certain extent, this is caused by the discovery of the accelerated expansion of the present Universe
at the end of the 1990s. One of the possible explanations of such acceleration is the assumption about
the existence in the Universe of a special form of matter~-- dark energy~\cite{AmenTsu2010}.
A distinctive feature of the latter is that it
must necessarily violate one of the energy conditions. For a convenient description of such a violation, one can introduce
some effective equation of state relating an effective pressure $p$ and energy density $\varepsilon$ of dark energy.
``Soft'' models of dark energy imply that it is described by matter violating the strong energy condition
when $p<-\varepsilon/3$. However, it is not impossible that, for an adequate description of the accelerated expansion of the Universe,
one has to involve a more exotic form of dark energy violating the null/weak energy condition when $p<-\varepsilon$.
Such exotic matter, filling the Universe homogeneously and isotropically on large scales,
supports the observable acceleration.

If the exotic matter does really exist in the Universe, a natural question arises as to the
possibility of the formation of localized wormhole-type objects supported by such matter. Indeed, the possibility of violating
the null/weak energy condition inherent in the exotic matter is just that ingredient which is necessary for ensuring the presence of
a nontrivial spacetime topology inherent in a wormhole.
In the simplest case, a description of the exotic matter can be given by using
the so-called ghost (or phantom) scalar fields, which may be massless~\cite{Bronnikov:1973fh,Ellis:1973yv,Ellis:1979bh}
or possess a potential energy~\cite{Kodama:1978dw,Kodama:1978zg}.
In turn, such static solutions can be generalized  to the case with rotation~\cite{Kashargin:2007mm,Kashargin:2008pk,Kleihaus:2014dla,Chew:2016epf,Kleihaus:2017kai,Chew:2019lsa}.

However, apart from the possibility of forming objects of the type pure wormholes  (i.e., systems containing only exotic matter and possessing a nontrivial  spacetime topology),
one can also imagine a situation where exotic matter is only one of components forming some mixed configurations containing other types of matter as well.
As an example of such mixed configurations with nontrivial topology, we can point out systems considered by us earlier, where, apart from a
 ghost scalar field, there is also either neutron matter~\cite{Dzhunushaliev:2011xx,Dzhunushaliev:2012ke,Dzhunushaliev:2013lna,Dzhunushaliev:2014mza,Aringazin:2014rva,Dzhunushaliev:2015sla,Dzhunushaliev:2016ylj},
 or an ordinary scalar field~\cite{Dzhunushaliev:2014bya}, or a chiral field~\cite{Charalampidis:2013ixa}.
  As in the case of pure wormholes, such mixed systems can also be generalized to the case where rotation is present~\cite{Hoffmann:2017vkf,Hoffmann:2018oml}.

In the present paper we focus attention on further studying mixed configurations supported by a ghost scalar field
(which ensures the presence of a nontrivial spacetime topology) and ordinary neutron star matter.
As mentioned above, we have earlier considered static neutron-star-plus wormhole configurations
of this type~\cite{Dzhunushaliev:2011xx,Dzhunushaliev:2012ke,Dzhunushaliev:2013lna,Dzhunushaliev:2014mza,Aringazin:2014rva,Dzhunushaliev:2015sla,Dzhunushaliev:2016ylj}.
Such mixed systems possess properties both of wormholes and of ordinary stars. Namely, on the one hand,
from the point of view of a distant observer, they are similar to ordinary neutron stars having masses and sizes which are typical for such stars.
On the other hand, the presence of a nontrivial wormholelike spacetime topology gives rise to the appearance of some new distinctive characteristics
which could, in principle, be revealed in astrophysical observations (for instance, this can be lensing effects~\cite{Dzhunushaliev:2014mza},
specific distributions of magnetic fields~\cite{Aringazin:2014rva},
 accretion disks and black hole mimickers~\cite{Dzhunushaliev:2016ylj}).
In principle, it is not impossible that, along with ordinary neutron stars, such systems might exist in nature,
since their physical characteristics are comparable to those typical of neutron stars.
However, since real neutron stars are rotating configurations, it is natural to generalize static mixed systems considered by us earlier
to the case of rotating systems. This is the goal of the present paper.

The paper is organized as follows. In Sec.~\ref{gen_eqs}, we state the problem and present
the general-relativistic equations for the systems under consideration.
These equations have been solved numerically, and the solutions are discussed in Sec.~\ref{num_sol} for static and rotating configurations. The question of their stability is briefly addressed in Sec.~\ref{stability}.
Finally, in Sec.~\ref{concl} we summarize
the results obtained.

\section{General equations}
\label{gen_eqs}

We consider a gravitating system consisting of a wormhole supported by a ghost scalar field $\phi$ and filled by
ordinary matter in the form of a neutron fluid. For simplicity,  we consider a massless scalar field.
The Lagrangian of this system can be chosen as follows:
\begin{equation}
\label{lagran_wh_star_poten}
L=-\frac{c^4}{16\pi G}R-\frac{1}{2}\partial_{\mu}\phi\partial^{\mu}\phi+L_{\text{fl}},
\end{equation}
with the curvature scalar $R$,
Newton's constant $G$, and the Lagrangian of the perfect isotropic fluid $L_{\text{fl}}=p$~\cite{Brown:1993}, where $p$ is the pressure of the fluid.

Varying the action with the Lagrangian~\eqref{lagran_wh_star_poten} with respect to the metric, we derive the Einstein equations
 \begin{equation}
\label{Ein_gen}
E_{\mu}^\nu	\equiv R_{\mu}^\nu - \frac{1}{2} \delta_{\mu }^\nu R -
	\frac{8\pi G}{c^4} T_{\mu }^\nu=0
\end{equation}
with the energy-momentum tensor
$$T_\mu^\nu=(\varepsilon+p)U_\mu U^\nu-\delta_\mu^\nu p-\partial_{\mu}\phi\partial^{\nu}\phi
+\frac{1}{2}\delta_\mu^\nu\partial_{\sigma}\phi\partial^{\sigma}\phi,
$$
where $\varepsilon$ is the energy density of the fluid, and $U_\mu$ represents the four-velocity of the fluid, with the norm of $U^\mu $ being a dimensionless quantity.

In order to describe rotating configurations, we use the following line element for a stationary, axially
symmetric spacetime~\cite{Kleihaus:2014dla,Chew:2016epf,Chew:2019lsa}:
\begin{equation}
\label{metric}
ds^2=e^f (d x^0)^2-e^{q-f}\left[e^b\left(d\eta^2+h d\theta^2\right)+h\sin^2\theta\left(d\varphi-\frac{\omega}{c}dx^0\right)^2\right],
\end{equation}
where the metric functions $f,q,b$, and $\omega$ depend solely on the radial coordinate $\eta$ and the polar angle $\theta$,  $x^0=c t$, and
the auxiliary function $h=\eta^2+\eta_0^2$ contains the throat parameter $\eta_0$.
The $z$-axis ($\theta=0$) represents the symmetry axis of the system.
Asymptotically (as $\eta\to \pm \infty$), the functions $f, q, b,\omega \to 0$; i.e., the spacetime approaches Minkowski spacetime.

In order to describe the neutron matter filling the wormhole, we have to choose an appropriate equation of state (EOS).
For the sake of simplicity, we employ here a polytropic equation of state which adequately approximates
a more or less realistic neutron-star EOS. Namely, we take 
\begin{equation}
\label{eqs_NS_WH}
p=K \rho_{b}^{1+1/n}, \quad \varepsilon = \rho_b c^2 +n p,
\end{equation}
with the constant $K=k c^2 (n_{b}^{(ch)} m_b)^{1-\gamma}$,
the polytropic index $n=1/(\gamma-1)$,
and $\rho_b=n_{b} m_b$ denotes the rest-mass density
of the neutron fluid. Here $n_{b}$ is the baryon number density,
$n_{b}^{(ch)}$ is a characteristic value of $n_{b}$,
$m_b$ is the baryon mass,
and $k$ and $\gamma$ are parameters
whose values depend on the properties of the matter the neutron star is composed of.

As in our previous works concerning mixed star-plus-wormhole systems
\cite{Dzhunushaliev:2012ke,Dzhunushaliev:2013lna,Dzhunushaliev:2014mza},
we here, for simplicity, take only one set of parameters for the neutron star fluid.
Namely, we choose
$m_b=1.66 \times 10^{-24}\, \text{g}$,
$n_{b}^{(ch)} = 0.1\, \text{fm}^{-3}$,
$k=0.0195$ and $\gamma=2.34$ \cite{Damour:1993hw},
adjusted to fit the equation of state II of Ref.~\cite{DiazAlonso:1985}.
We employ these values of the parameters in the numerical calculations of Sec.~\ref{num_sol}.

We study here uniform (or rigid) rotation of the neutron star fluid; such an assumption is well justified for most neutron stars~\cite{Haensel:2007yy}.
In this case the four-velocity is of the form
\begin{equation}
\label{four_vel}
U^\mu=\left(u,0,0,\frac{\Omega}{c}u\right),
\end{equation}
where $\Omega$ is a constant parameter denoting the angular velocity of the system.

For numerical calculations, it is convenient to introduce the dimensionless variables
\begin{equation}
\label{dmls_var}
\tilde\eta=\frac{\eta}{L},
\quad \tilde\phi=\frac{\sqrt{8\pi G}}{c^2}\,\phi, \quad \tilde\omega=\frac{L}{c}\omega, \quad \tilde\Omega=\frac{L}{c}\Omega,
\quad \text{where} \quad L=\frac{c^2}{\sqrt{8\pi G}\phi_1},
\end{equation}
and to use the standard reparametrization of the fluid density,
\begin{equation}
\label{theta_def}
\rho_b=\rho_{b 0} \Theta^n~,
\end{equation}
where $\rho_{b 0}$ is some characteristic  density of the neutron star fluid. The parameter $\phi_1$ appearing in
\eqref{dmls_var} corresponds to the central value of the derivative of the scalar field with respect to the radial coordinate
(the square of this derivative being the ``kinetic energy'' of the scalar field).

Next, using the fact that for the four-velocity $U^\mu U_\mu=1$
and making use of the metric \eqref{metric} and the expression \eqref{four_vel}, the velocity $u$
can be expressed in terms of the metric functions $f, q$, and $\omega$ as
\begin{equation}
\label{expres_u}
u^2=e^{-f}\left[1-e^{q-2 f}\left(\eta^2+\eta_0^2\right) \sin^2\theta\left(\Omega-\omega\right)^2\right]^{-1}.
\end{equation}
This expression is already written in terms
of the dimensionless variables~\eqref{dmls_var}; in order to make the notation simpler,  hereafter we usually omit the tilde sign over the dimensionless variables.

In turn, the conservation law $\nabla_{\mu}T^{\mu\nu}=0$ gives the differential equations
$$\frac{\partial_r p}{\varepsilon+p}=\frac{\partial_r u}{u},\quad \frac{\partial_\theta p}{\varepsilon+p}=\frac{\partial_\theta u}{u}.$$
Substituting here the EOS~\eqref{eqs_NS_WH} [using also Eq.~\eqref{theta_def}] and integrating these equations, one can find the following expression for the function
$\Theta$ in terms of the four-velocity:
$$\Theta=c_0 u-\frac{1}{\sigma (n+1)},$$
where $c_0$ is an integration constant and $\sigma=K \rho_{b0}^{1/n}/c^2$.
Introducing instead of $c_0$ a new arbitrary constant $A=\left[c_0\sigma (n+1)\right]^{-1}$,
the above expression can be rewritten as
\begin{equation}
\label{expres_Theta}
\Theta=\frac{u-A}{A\sigma(n+1)}.
\end{equation}
The physical meaning of $A$ is that it is the time component of the four-velocity of the fluid at the edge of the neutron fluid where
$\Theta$ vanishes (see below).

 The Einstein equations \eqref{Ein_gen} give the following system of equations for the metric functions $f, q, b,$ and $\omega$
[written in terms of the dimensionless variables~\eqref{dmls_var}]:
\begin{align}
&f_{\eta\eta}-e^{-2 f+q} \sin^2\theta \left(h\omega_\eta^2+\omega_\theta^2\right)+\left(\frac{2\eta}{h}+\frac{1}{2}q_\eta\right)f_\eta+\frac{f_{\theta\theta}}{h}
+\frac{1}{2h}\left(2\cot\theta+q_\theta\right)f_\theta\nonumber\\
&=B e^{b- f+q}\Theta^n\left\{
e^f u^2+\sigma\left[2+e^f\left(n+1\right)u^2\right]\Theta+e^{q-f}u^2 h \sin^2\theta\left[1+\sigma\left(n+1\right)\Theta\right]
\left(\Omega-\omega\right)^2
\right\},
\label{eq_f}\\
&q_{\eta\eta}+\frac{1}{2}q_\eta^2+\frac{3\eta}{h}q_\eta+\frac{1}{h}\left(q_{\theta\theta}+\frac{1}{2}q_\theta^2+2\cot\theta q_\theta
\right)=4 B \sigma e^{b- f+q}\Theta^{n+1},
\label{eq_q}\\
&b_{\eta\eta}-\frac{3}{2}e^{-2 f+q} \sin^2\theta \left(h \omega_\eta^2+ \omega_\theta^2\right)+\frac{1}{2}\left(f_\eta^2-q_\eta^2\right)
+\frac{1}{h}\left[b_{\theta\theta}+\eta \left(b_\eta-2 q_\eta\right)+\frac{1}{2}\left(f_\theta^2-q_\theta^2\right)-
2\cot\theta q_\theta-\frac{2\eta^2}{h}+2
\right]\nonumber\\
&=2 B e^{b- f+q}\Theta^n\left\{-\sigma\Theta+e^{q-f}u^2 h\sin^2\theta\left[1+\sigma\left(n+1\right)\Theta\right]\left(\Omega-\omega\right)^2
\right\}+\phi_\eta^2+\frac{\phi_\theta^2}{h},
\label{eq_b}\\
&\omega_{\eta\eta}+\left(\frac{4\eta}{h}-2 f_\eta+\frac{3}{2}q_\eta\right)\omega_\eta+
\frac{1}{h}\left[\omega_{\theta\theta}+\left(3\cot\theta-2 f_\theta+\frac{3}{2}q_\theta\right)\omega_\theta
\right]=-2 B e^{b+q}u^2\Theta^n\left[1+\sigma\left(n+1\right)\Theta\right]\left(\Omega-\omega\right).
\label{eq_omega}
\end{align}
Here the lower indices denote differentiation with respect to the corresponding coordinate and $B=\rho_{b0}c^2/\phi_1^2$.
These equations represent the following combinations of the components of the Einstein equations~\eqref{Ein_gen}:
$\big(E_t^t-E_\eta^\eta-E_\theta^\theta-E_\varphi^\varphi+2\,\omega/c E_\varphi^t\big)=0$,
$\left(E_\eta^\eta+E_\theta^\theta\right)=0$, $\left(E_\varphi^\varphi-E_\eta^\eta-E_\theta^\theta-\omega/c E_\varphi^t\right)=0$,
and $E_\varphi^t=0$.

In turn, the equation for the scalar field $\nabla_\mu\nabla^\mu\phi=0$ yields
\begin{equation}
\phi_{\eta\eta}+\frac{\phi_{\theta\theta}}{h}+\left(\frac{2\eta}{h}+\frac{1}{2}q_\eta\right)\phi_\eta+
\frac{1}{h}\left(\cot\theta+\frac{1}{2}q_\theta\right)\phi_\theta=0.
\label{eq_phi}
\end{equation}

In addition to the Einstein equations~\eqref{eq_f}-\eqref{eq_omega}, which are elliptic partial differential equations, one has two more equations of gravitation $E^\eta_\theta=0$ and
$\left(E^\eta_\eta-E^\theta_\theta\right)=0$, whose structure is not of Laplace form; they may be regarded as ``constraints.''
According to arguments given in Ref.~\cite{Wiseman:2002zc}, in order to find a self-consistent solution to
Eqs.~\eqref{eq_f}-\eqref{eq_omega}, it is sufficient to verify that the constraint equation $E^\eta_\theta=0$ is satisfied on all boundaries;
this in turn implies that this constraint is also satisfied throughout the region of integration.
Then the constraint $\left(E^\eta_\eta-E^\theta_\theta\right)=0$ must be imposed only at a single point to ensure that it is satisfied throughout the integration region as well.
In our numerical calculations we will always verify that these constraints are satisfied, in order to have self-consistent numerical solutions.

\section{Numerical solutions}
\label{num_sol}
In this section we solve Eqs.~\eqref{eq_f}-\eqref{eq_phi} to get solutions describing static and rotating configurations with system parameters
lying in the physically relevant range.

\subsection{Boundary conditions}

Our aim is to find globally regular solutions describing localized, finite-mass configurations embedded in an asymptotically flat spacetime.
To do this, one has to impose appropriate boundary conditions for the metric functions at the center ($\eta=0$), at spatial infinity ($\eta\to \infty$),
on the positive $z$-axis ($\theta=0$), and, using reflection symmetry with respect to $\theta\to \pi-\theta$,
in the equatorial plane ($\theta=\pi/2$). Namely, we take
\begin{align}
\label{BCs}
\begin{split}
&\left. \frac{\partial f}{\partial \eta}\right|_{\eta = 0} =
\left. \frac{\partial q}{\partial \eta}\right|_{\eta = 0} = \left. \frac{\partial b}{\partial \eta}\right|_{\eta = 0} =
\left. \frac{\partial \omega}{\partial \eta}\right|_{\eta = 0} =  0,
\left. \phi \right|_{\eta = 0} =0;\\
&\left. f \right|_{\eta = \infty} = \left. q \right|_{\eta = \infty} =\left. b \right|_{\eta = \infty} =
	\left. \omega \right|_{\eta = \infty} = 0, \left. \phi \right|_{\eta = \infty} =\text{const.}  ; \\
&\left. \frac{\partial f}{\partial \theta}\right|_{\theta = 0,\pi/2,\pi} =
\left. \frac{\partial q}{\partial \theta}\right|_{\theta = 0,\pi/2,\pi} =
\left. \frac{\partial b}{\partial \theta}\right|_{\theta = 0,\pi/2,\pi} =
	\left. \frac{\partial \omega}{\partial \theta}\right|_{\theta = 0,\pi/2,\pi} =
\left. \frac{\partial \phi}{\partial \theta}\right|_{\theta = 0,\pi/2,\pi} =0.
\end{split}
\end{align}
Note here that, in order to ensure the absence of a conical singularity, we must take $b|_{\theta=0,\pi}=0$ (the elementary flatness condition).

\subsection{Asymptotic behavior}

Asymptotic flatness of the spacetime implies that the metric approaches the Minkowski
metric at spatial infinity, i.e.,  $f,q,b, \omega\to 0$ and $\phi\to \text{const}$, asymptotically.
For the extraction of the global charges, one needs to study the
behavior of the metric functions at infinity,
$$
f\approx-\frac{2 G M}{c^2 \eta},\quad \omega\approx \frac{2G J}{c^2 \eta^3}.
$$
The total mass of the star $M$ and the angular momentum $J$ appearing in the above expressions may then be represented
 in the form~\cite{Chew:2019lsa}
\begin{equation}
\label{expres_mass_mom}
M=\frac{c^2}{2G}\lim_{\eta\to\infty}\eta^2\partial_\eta f, \quad J=\frac{c^2}{2G}\lim_{\eta\to\infty}\eta^3\omega.
\end{equation}

\subsection{Central region and surface of the configurations}

The central  energy density and pressure of the system under consideration can be obtained from Eqs.~\eqref{eqs_NS_WH} and
\eqref{theta_def} using  Eqs.~\eqref{expres_Theta} and \eqref{expres_u} evaluated at the center (in the dimensional form):
$$
P_c=\rho_{b0} c^2 \sigma\Theta_c^{n+1}, \quad \varepsilon_c=\rho_{b0}c^2\Theta_c^{n}\left(1+n\sigma \Theta_c\right)
\quad \text{with} \quad \Theta_c=\frac{u_c-A}{A\sigma(n+1)}.
$$
By choosing different values of the arbitrary constant $A$, it is possible
to construct sequences of solutions describing configurations possessing different physical characteristics (see below).

The configuration under consideration is a system in which 
neutron star matter 
is coupled to a ghost scalar field.
Formally, the energy density of the latter vanishes only at infinity,
whereas the neutron star has a finite boundary and a surface
with radius $\eta_b(\theta)$ where $\varepsilon\to 0$ (correspondingly, $p \to 0$ as well).
One can  construct an embedding diagram that describes the intrinsic geometry of the stellar surface [i.e., the spacetime slice with $t=\text{const.}$ and $\eta=\eta_b(\theta)$].
To do so, it is necessary to embed the stellar surface in a flat three-dimensional space~\cite{Friedman:1986tx}.
Then the metric of the star's surface induced by the four-dimensional metric \eqref{metric} is
$$
ds_b^2=e^{q_b-f_b}\left\{e^{b_b}\left[\left(\partial_\theta \eta_b\right)^2+h_b
\right]d\theta^2+h_b\sin^2\theta d\varphi^2
\right\},
$$
and here the index $b$ corresponds to the boundary of the neutron star where the metric functions are given [for example, $f_b\equiv f\left(\eta_b(\theta),\theta\right)$].
If we change to cylindrical coordinates $\{\rho, z, \varphi\}$ for the flat space, we then have the following formulas~\cite{Friedman:1986tx}:
$$\rho(\theta)=e^{\left(q_b-f_b\right)/2} \sqrt{h_b} \sin\theta, \quad
z(\theta)=\int_{\theta}^{\pi/2}d\theta^\prime
\left\{e^{q_b-f_b+b_b}\left[\left(\frac{d \eta_b}{d\theta^\prime}\right)^2+h_b\right]-\left(\frac{d\rho}{d\theta^\prime}\right)^2
\right\}^{1/2},$$
where $h_b=\eta_b^2+\eta_0^2$.

The equatorial and polar radii of the embedded surface are then given by
\begin{align}
&R_e=\rho\left(\theta=\frac{\pi}{2}\right)=\left. e^{\left(q_b-f_b\right)/2}\sqrt{h_b}\right|_{\theta = \pi/2} ,
\label{rad_eq}
\\
&R_p=z(\theta=0)=\int_{0}^{\pi/2}d\theta^\prime
\left\{e^{q_b-f_b+b_b}\left[\left(\frac{d \eta_b}{d\theta^\prime}\right)^2+h_b\right]-\left(\frac{d\rho}{d\theta^\prime}\right)^2
\right\}^{1/2}.
\label{rad_pol}
\end{align}

Let us now find an expression for the limiting angular velocity of rotation (or the Keplerian limit) of the configuration under consideration.
This limit corresponds to the case of a system rotating with the angular velocity $\Omega$ approaching
the angular velocity $\Omega_p$ of a free particle moving on a circular orbit in the equatorial plane.
In this case, gravitational forces are 
no longer sufficiently strong as to keep such a particle confined
 to the stellar surface, and this ultimately results in losing mass.

In order to derive the corresponding expression for $\Omega_p$, we use the geodesic equation for the angular
velocity of the particle~\cite{Friedman:2013xza}
\begin{equation}
\label{expres_Op}
\left(\Omega_p-\omega_b\right)^2-2 a_p c\left(\Omega_p-\omega_b\right)+c^2 b_p=0,
\end{equation}
where $\omega_b\equiv \omega(\eta_b,\pi/2)$ and
$$
a_p= \left.\frac{1}{c}\frac{\partial_\eta \omega}{\partial_\eta (q-f)+2\eta/h}\right|_{\eta_b,\pi/2} ,\quad
b_p= \left. -\frac{e^{2 f-q}\partial_\eta f}{h\left[\partial_\eta (q-f)+2\eta/h\right]}\right|_{\eta_b,\pi/2} .
$$
Solving Eq.~\eqref{expres_Op} for $\Omega_p$, we have
$$\frac{\Omega_p}{c}=\frac{\omega_b}{c}+a_p+\sqrt{a_p^2-b_p}.$$
By definition, the Keplerian angular velocity is $\Omega_K=\Omega_p$. A necessary condition for
a stationary rotating object to exist is that the equatorial velocity of a fluid element is smaller than the Keplerian velocity
of a free particle moving on a circular orbit in the equatorial plane. The Keplerian velocity is related to the Keplerian angular velocity $\Omega_K$,
also called the mass-shedding angular velocity.

\subsection{Numerical approach}

We have solved numerically the system of five coupled nonlinear elliptic partial differential equations~\eqref{eq_f}-\eqref{eq_phi} for the functions $f, q, b,\omega$, and $\phi$
together with the  boundary conditions~\eqref{BCs}. Bearing in mind that we are interested in even parity solutions symmetrical about the equatorial plane $\theta=\pi/2$,
all calculations have been performed only in the region $0\leq \theta \leq \pi/2$.
Furthermore, for the numerical computations, 
we have introduced the new compactified coordinate
\begin{equation}
	\bar \eta=\frac{\eta}{1+\eta} ,
	\label{comp_coord}
\end{equation}
in order to map the infinite interval $[0,\infty)$ to the finite region $[0,1]$.

The results of the numerical computations for axisymmetric systems presented in Sec.~\ref{st_rot_sols} have been obtained using the package FIDISOL~\cite{Schoenauer:1989}.
This package employs the Newton-Raphson method which provides  an iterative procedure for obtaining
an exact solution starting from some approximate solution (an initial guess). As the initial guess, it is possible to use spherically symmetric solutions
describing nonrotating (i.e., static) configurations, which can be obtained by solving Eqs.~\eqref{eq_f}-\eqref{eq_phi} with $\Omega, \omega, b=0$.
In this limit the equations reduce to the system of three ordinary differential equations for the functions $f,q$, and $\phi$.
This latter set of equations has been solved using the package {\it Mathematica}.

The partial differential equations~\eqref{eq_f}-\eqref{eq_phi} have been solved on a grid of $361\times 61$ points which
covers  the integration region  $0\leq \bar \eta \leq 1$ [given by the compactified radial coordinate, Eq.~\eqref{comp_coord}] and $0\leq \theta \leq \pi/2$.
In solving these equations, we have kept track of the behavior of the function $\Theta$, defined in Eq.~\eqref{theta_def}, at every point in space $(\eta, \theta)$.
This enables us to determine the location of the edge of the neutron fluid $\eta_b(\theta)$ where $ \Theta \to 0$.
If $ \Theta < 0$, this indicates that the meshpoint is outside the fluid; in this case we set to zero the neutron matter parts in the
right-hand sides of Eqs.~\eqref{eq_f}-\eqref{eq_omega}, retaining only the terms coming from the scalar field
(that is, we solved the 
Einstein-scalar equations).

\subsection{Static and rotating solutions}
\label{st_rot_sols}

In this subsection we discuss the families of solutions describing the systems under consideration, obtained by using the numerical approach outlined above.
In particular, we employed 
the following strategy: As seed configurations, we first obtained nonrotating, spherically symmetric configurations;
to do this, we started the integration in the vicinity of the center using the Taylor series expansion for the metric functions
$$
f\approx f_c+\frac{1}{2}f_2 \eta^2+ \ldots, \quad q\approx q_c+\frac{1}{2}q_2 \eta^2+ \ldots
$$
In order to ensure asymptotic flatness of the spacetime,
we had to take 
central values $f_c$ and $q_c$
that provide $f, q \to 0$ at spatial infinity (i.e., we dealt with
an eigenvalue problem). Then, by varying the integration constant $A$ appearing in the expression
\eqref{expres_Theta}, we could obtain a family of nonrotating configurations ($\omega=\Omega=0$) 
with varying masses and radii of the neutron star fluid;
these systems are parametrized by $A$.

\begin{figure}[t]
\includegraphics[width=1.\linewidth]{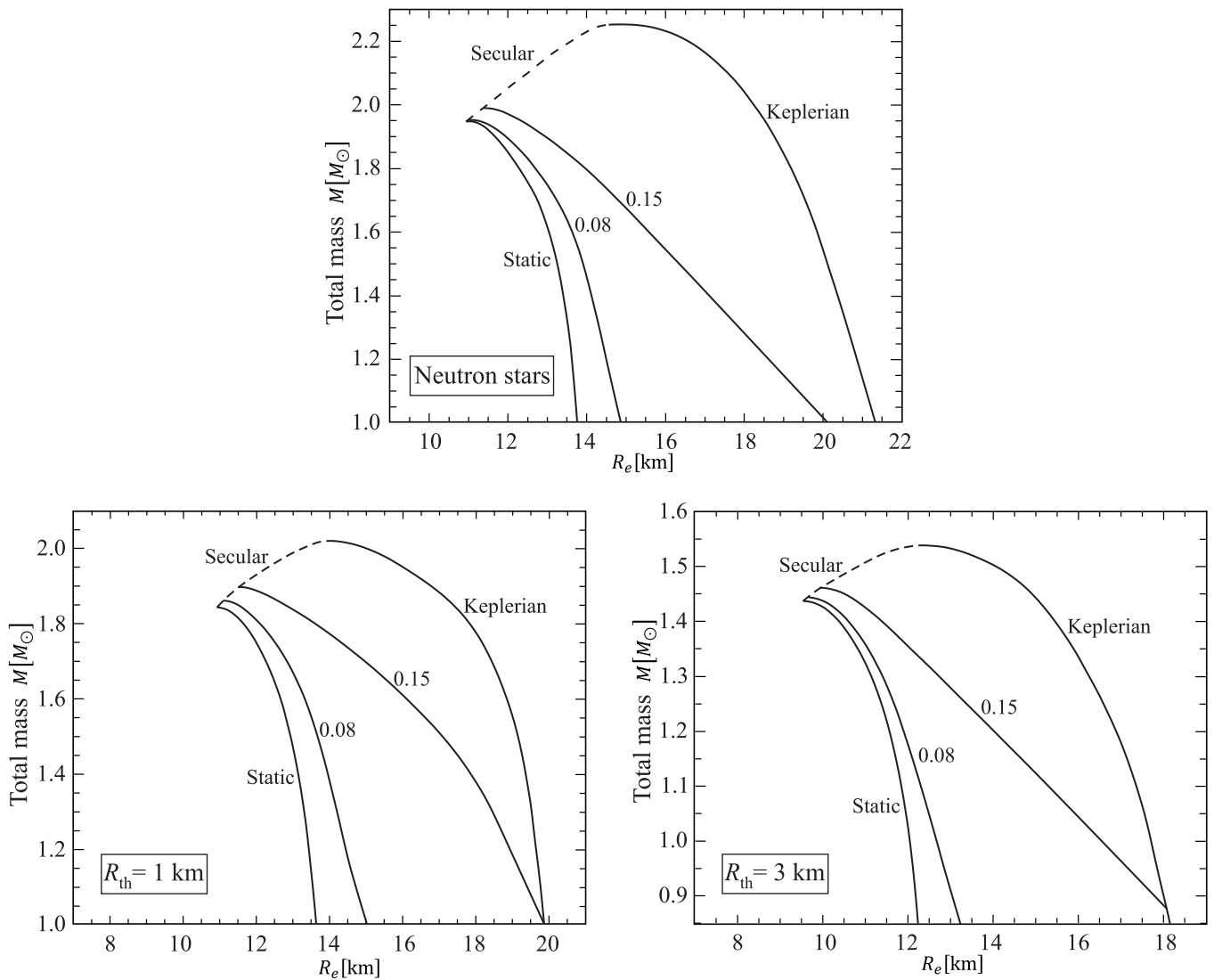}
\caption{
The mass-radius relation for configurations rotating with different
angular velocities $\Omega$ (designated by the numbers near the curves) in the physically relevant domain.
The top panel shows the curves for pure neutron stars.
The curves for mixed neutron-star-plus-wormhole systems are constructed for the
fixed values of the throat radius $R_{\text{th}}=1\, \text{km}$ and $R_{\text{th}}=3 \,\text{km}$ [for its definition, see below Eq.~\eqref{radius_equat}].
For all configurations, we took $L=10\, \text{km}$
 and the parameters of the polytropic fluid given below Eq.~\eqref{eqs_NS_WH}.
The dimensionless angular velocity $\Omega=0.01$ corresponds to a frequency $f\approx 47.7 \,\text{Hz}$.
For $\Omega=0.15$, the frequency is $f\approx 716 \,\text{Hz}$ (the frequency of the fast rotating pulsar~\cite{Hessels:2006ze}).
}
\label{fig_M_R_eq}
\end{figure}

Next, using the nonrotating configurations obtained as an initial guess, we slowly increased the angular velocity ($\Omega\neq 0$) keeping track of changes in the characteristics
of the configurations under investigation. The results of these computations are exhibited in Fig.~\ref{fig_M_R_eq} which shows the mass-radius relation for the objects rotating
with different angular velocities.
To plot these curves, the following  dimensional quantities have been employed
 [they are obtained using Eqs.~\eqref{dmls_var} and \eqref{expres_mass_mom}]:
$$\eta_b(\theta)=\frac{\bar \eta_b}{1-\bar \eta_b} L, \quad
M=\frac{c^2 L}{2 G}\lim_{\bar \eta\to 1}\bar \eta^2\frac{\partial f}{\partial \bar \eta}.
$$
To determine the equatorial radius $R_e$, it is necessary to substitute this expression for $\eta_b(\theta)$ into Eq.~\eqref{rad_eq}.
In turn, the spin frequency of the systems under consideration is determined as
$$
f\equiv\frac{\Omega}{2\pi}=\frac{c}{2\pi L} \tilde\Omega.
$$

Examples of the domains of the configurations are shown in Fig.~\ref{fig_M_R_eq}.
Their boundaries arise according to the following physical reasons:
\begin{itemize}
\item[(i)] 
The left boundary of the domain corresponds to the set of nonrotating systems ($\Omega=0$); their mass increases monotonically with decreasing the radius 
up to the maximal mass.
In the case of pure neutron stars (i.e., the systems without a wormhole) the presence of
this maximum 
signals the transition of the neutron stars 
from stable to unstable ones, where the radial oscillations acquire an unstable mode \cite{Harrison:1965,Kokkotas:2000up}. However, for the mixed configurations under consideration, 
all static configurations are unstable, since there is always an unstable radial mode present in the mixed configurations that is inherited from the static wormhole~\cite{Dzhunushaliev:2013lna}.
However, at the maximum mass, a second unstable radial mode is expected to arise, that would now result from the instability of the neutron star matter with respect to radial oscillations.
\item[(ii)] The right boundary of the domain is the locus  of the points corresponding to configurations rotating with the Keplerian angular velocity $\Omega=\Omega_K$.
Here, with decreasing radius, 
the mass of the system also increases monotonically up to a maximal value that, as in the case of the pure neutron stars, exceeds the maximum mass of the static configurations considerably.
\item[(iii)] The two boundaries discussed in items (i) and (ii) are connected
by the so-called secular instability line (shown by the dashed line) which is the locus of
the extreme points corresponding to maxima of masses of the systems possessing different constant angular velocities $\Omega$.
For the pure neutron stars, this secular instability line again separates the unstable (located to the left of this line) and stable
configurations. We conjecture the presence of such a secular instability also for the mixed configurations (see below in Sec.~\ref{stability}).
\end{itemize}

To obtain the curves corresponding to the configurations rotating with the Keplerian angular velocity, we successively increased the value of
$\Omega$  up to the point where $\Omega\approx\Omega_p$. Technically, it is not possible to continue numerical integration to the point $\Omega=\Omega_p$;
for this reason, to obtain physical parameters of the objects rotating at the Keplerian limit,
we extrapolated them as functions of $\Omega$ in the limit $\Omega\to\Omega_p$~\cite{Kleihaus:2016dui}.

A crucial ingredient of the systems under consideration is the wormhole throat, which is defined as a surface possessing minimum area.
To describe it, consider the circumferential radius $R_e(\eta)$ in the equatorial plane,
\begin{equation}
\label{radius_equat}
 R_e(\eta)=\left. \sqrt{h}e^{(q-f)/2}\right|_{\theta=\pi/2}.
\end{equation}
When $\eta\to 0$, the circumferential radius reaches its minimum value $R_{\text{th}}\equiv  \left. R_e(0)\right|_{\theta=\pi/2}$,
and the corresponding surface possessing minimum area corresponds to the wormhole throat.

Figure~\ref{fig_M_R_eq} shows the results of the calculations for ordinary neutron stars (i.e., for systems without a wormhole)
and for mixed neutron-star-plus-wormhole configurations with two fixed values of the throat radius $R_{\text{th}}$.
By comparing the behavior of the mass-radius curves of the systems under consideration, one can see that the inclusion of a wormhole at the center of a neutron star
does not change the qualitative behavior of the boundary curves and thus the domains. 
As the size of the throat increases, the maximum masses of the systems
under consideration decrease together with their sizes.
Whereas the radii remain comparable to those of ordinary neutron stars, the masses fall short of reaching the observational bound.

\subsection{Stability}
\label{stability}

Let us now address the question of stability of the systems under consideration in more detail.
In Refs.~\cite{Dzhunushaliev:2013lna,Dzhunushaliev:2014mza}, it
was demonstrated that spherically symmetric mixed neutron-star-plus-wormhole systems are unstable with respect to radial linear perturbations. 
As in the case of static Ellis wormholes supported by a massless ghost scalar field~\cite{Gonzalez:2008wd,Gonzalez:2008xk,Blazquez-Salcedo:2018ipc}
and of wormholes consisting of scalar fields with a potential~\cite{Dzhunushaliev:2013lna,Dzhunushaliev:2017syc},
this radial instability is caused by the presence of a ghost scalar field in the static mixed configurations. 
Inclusion of rotation tremendously complicates a linear stability analysis in four spacetime dimensions.
For instance,  a linear mode analysis without approximations was performed only recently for rapidly rotating neutron stars  \cite{Kruger:2019zuz,Kruger:2021zta}.
It presents a challenge to extend such a type of analysis to mixed neutron-star-plus-wormhole systems.
We note, however, that in
the case of five spacetime dimensions the presence of rotation does lead to the disappearance
of the pertinent radial instability of Ellis wormholes for sufficiently rapid rotation~\cite{Dzhunushaliev:2013jja}.
Thus this instability might also disappear for mixed systems when rotating sufficiently fast.

However, even without performing an ambitious stability analysis, it is 
clear that, for stable configurations,
as the mass increases, their radii should decrease~\cite{Zeldovich:1996}.
This fact forms the basis of the turning-point method~\cite{Friedman:1988er},
which does not require a 
perturbative stability analysis. In the case of static systems, this method permits one to estimate stability by
studying the behavior of a mass-central density curve $M(\varepsilon_c)$.
According to this method, instability occurs
at the first point where $\partial M/\partial\varepsilon_c=0$ (the turning point).
Such an instability evolves on a secular timescale~
\cite{Friedman:1988er}.

For rotating configurations, apart from the central density, there are extra parameters like the angular velocity and the angular momentum.
It is then possible to construct a sequence of rotating configurations for which the central energy density varies in a range that is constrained by stability limits when the selected
second parameter is held fixed~\cite{Cipolletta:2015nga}.
As demonstrated in Ref.~\cite{Friedman:1988er}, such a turning-point method can also be employed to study the stability of uniformly rotating stars.
For instance, for a family of configurations rotating with a constant angular velocity $\Omega$,
the boundary between secularly stable and unstable systems is located at the turning point where $\left.\partial M(\varepsilon_c, \Omega)/\partial \varepsilon_c\right|_{\Omega=\text{const.}}=0$.
Thus, by finding the turning points for configurations rotating with different fixed values of $\Omega$ and connecting them, one can construct the secular instability line.
For pure neutron stars, the secular instability line is depicted by the dashed curve in the top panel of Fig.~\ref{fig_M_R_eq}.
The point where this line intersects the Keplerian sequence corresponds to a system possessing the largest possible angular velocity.

In the case of the mixed neutron-star-plus-wormhole systems under consideration, this simple stability criterion is also applied for the secular instability.
Thus, analogously to rapidly rotating ordinary neutron stars, we have connected the maxima of the mass-radius curves by a secular line in Fig.~\ref{fig_M_R_eq}
to 
exclude certainly unstable systems located to the left of this line.

However, as already pointed out above, the static spherically symmetric mixed neutron-star-plus-wormhole systems 
possess an unstable radial mode due to the ghost scalar field~\cite{Dzhunushaliev:2013lna,Dzhunushaliev:2014mza}.
By continuity, it is therefore expected that the rotating axisymmetric systems studied in the present paper
will be dynamically unstable as well, at least for slow rotation. In contrast, for rapid rotation, it is conceivable that this instability might disappear,
analogously to the case of rotating wormholes in five spacetime dimensions~\cite{Dzhunushaliev:2013jja}.
Resolution of this question will require a mode analysis of these rapidly rotating mixed systems.

\section{Conclusion}
\label{concl}

In the present paper, we have continued our previous
investigations of the mixed neutron-star-plus-wormhole configurations
began in Refs.~\cite{Dzhunushaliev:2011xx,Dzhunushaliev:2012ke,Dzhunushaliev:2013lna,Dzhunushaliev:2014mza,Aringazin:2014rva,Dzhunushaliev:2015sla,Dzhunushaliev:2016ylj}.
For this purpose, we have generalized the static systems considered earlier by including rapid rotation
and studied its influence on the physical characteristics of such mixed objects. In order to have a nontrivial wormhole-type spacetime topology in the system,
we have used a massless ghost scalar field. In turn, the neutron matter has been modeled by the simplest polytropic equation of state in the form~\eqref{eqs_NS_WH}
that adequately approximates more realistic EOSs of matter at nuclear density.
The use of such a polytropic EOS enabled us to integrate analytically the differential equations coming from the conservation law
and to represent expressions for the neutron fluid energy density and pressure in terms of metric functions.
Together with the ghost scalar field, these expressions serve as a source of a gravitational field modeled within Einstein's general relativity.

We now summarize the results obtained:
\begin{itemize}
\item[(i)] We have found regular solutions to the Einstein equations sourced by a massless ghost scalar field and neutron star matter. These solutions describe families
of static and uniformly rotating mixed neutron-star-plus-wormhole configurations possessing nontrivial spacetime topology.
\item[(ii)] By choosing two different fixed values of the throat radius, the mass-radius relations for the systems under consideration have been constructed.
This enabled us to establish the physically relevant domain which is bounded by the Keplerian limit and the secular instability line.
\item[(iii)] It has been demonstrated that the physical characteristics of the rotating mixed neutron-star-plus-wormhole configurations
are close to those typical of rotating neutron stars.
\end{itemize}

Also, the question of stability has been discussed. It was pointed out that the rotating mixed systems are expected to be dynamically unstable,  at least for slow rotation,
but one might expect that a rapid rotation could stabilize them, as it takes place in the case of rotating wormholes in five spacetime dimensions.

\section*{Acknowledgments}

This research has been funded by the Science Committee of the Ministry of Science and Higher Education of the Republic of Kazakhstan
(Grant  No.~AP14869140, ``The study of QCD effects in non-QCD theories'').
We are also grateful to the DFG RTG 1620 \textit{Models of Gravity} and the Research Group Linkage Programme of the Alexander von Humboldt Foundation for the support of this research.
 V.D. and V.F.  would like to thank the Carl von Ossietzky University of Oldenburg for hospitality while this work was carried out.

\end{document}